\documentclass[sigconf]{acmart} %

\usepackage{tabularx}

\usepackage{makecell}

\usepackage{array}
\newcolumntype{L}[1]{>{\raggedright\let\newline\\\arraybackslash\hspace{0pt}}m{#1}}
\newcolumntype{C}[1]{>{\centering\let\newline\\\arraybackslash\hspace{0pt}}m{#1}}
\newcolumntype{R}[1]{>{\raggedleft\let\newline\\\arraybackslash\hspace{0pt}}m{#1}}

\AtBeginDocument{%
  \providecommand\BibTeX{{%
    \normalfont B\kern-0.5em{\scshape i\kern-0.25em b}\kern-0.8em\TeX}}}

\setcopyright{acmcopyright}
\copyrightyear{2022}
\acmYear{2022}
\acmDOI{}

\acmConference[EASE '22]{EASE22: Evaluation and Assessment in Software Engineering 2022}{June 13-14, 2022}{Gothenburg, Sweden}

\begin{document}

\title[Text and Team: What Article Metadata Characteristics Drive Citations in Software Engineering?]{Text and Team: What Article Metadata Characteristics Drive Citations in Software Engineering?}

\author{Lorenz Graf-Vlachy}
\email{{lorenz.graf-vlachy|daniel.graziotin|stefan.wagner}@iste.uni-stuttgart.de}
\orcid{0000-0002-0545-6643}
\additionalaffiliation{%
  \institution{TU Dortmund University}
  \city{Dortmund}
  \state{Germany}
}
\author{Daniel Graziotin}
\orcid{0000-0002-9107-7681}
\author{Stefan Wagner}
\orcid{0000-0002-5256-8429}
\affiliation{%
  \institution{University of Stuttgart, Institute of Software Engineering}
  \streetaddress{Universit{\"a}tsstraße 38}
  \city{Stuttgart}
  \country{Germany}
  \postcode{70569}
}

\begin{abstract}
  \textit{Context}: Citations are a key measure of scientific performance in most fields, including software engineering. However, there is limited research that studies which characteristics of articles' metadata (title, abstract, keywords, and author list) are driving citations in this field.
  \textit{Objective}: In this study, we propose a simple theoretical model for how citations come to be with respect to article metadata, we hypothesize theoretical linkages between metadata characteristics and citations of articles, and we empirically test these hypotheses.
  \textit{Method}: We use multiple regression analyses to examine a data set comprising the titles, abstracts, keywords, and authors of 16,131 software engineering articles published between 1990 and 2020 in 20 highly influential software engineering venues.
  \textit{Results}: We find that number of authors, number of keywords, number of question marks and dividers in the title, number of acronyms, abstract length, abstract propositional idea density, and corresponding authors in the core Anglosphere are significantly related to citations.
  \textit{Conclusion}: Various characteristics of articles' metadata are linked to the frequency with which the corresponding articles are cited. These results partially confirm and partially go counter to prior findings in software engineering and other disciplines.
\end{abstract}



\maketitle

\section{Introduction}

Citations are probably the most important metric used in science to assess researchers' performance~\cite{Cronin.1994, Aksnes.2009, Adler.2009}. Despite mounting and substantial criticism of citations as an evaluation criterion for both scientific venues and individual researchers~\cite{Seeber.2019, Letrud.2019, Adler.2009}, they (or related concepts like the h-index~\cite{Hirsch.2005}) are widely used to make important assessments and subsequently decisions about researchers and even entire research groups~\cite{vanRaan.2006}. Specifically, citations are frequently considered in hiring and promotion decisions~\cite{Moher.2018}. More generally, citations are widely viewed as a metric that reflects a scientist's impact on their field~\cite{Aksnes.2003, Moed.2005}, as is evident, for instance, in the prominent display of citation counts at the top of every Google Scholar profile.

Consequently, prior researchers have sought to determine what drives citations. This debate takes place in various disciplinary fields, including most notably in the meta-scientific fields of bibliometrics and scientometrics~\cite{Tahamtan.2016, Onodera.2021}. Importantly, several authors identified characteristics of an article's metadata as drivers of citations. For instance, researchers studied the impact of the length of an article's abstract and title on citations across various disciplines~\cite{vanWesel.2014}. Similarly, researchers have identified the number of authors as being linked to citations~\cite{Aksnes.2003, vanWesel.2014}.

However, there is a dearth of empirical research in software engineering that systematically develops hypotheses on and tests which of such characteristics drive citations in this field. While some groundbreaking and promising work has been done, it is limited in several ways. Some work does not relate to articles' metadata but studies the effect of article content on citations~\cite{Molleri.2018}. Other work is mostly descriptive in nature and studies only a small set of drivers~\cite{Garousi.2016}. The arguably most extensive exemplar of extant work on the drivers of citations specifically in software engineering is a non-peer-reviewed working paper that studies journal but not conference papers~\cite{Mantyla.2019}.

This limited scholarship is regrettable because just as prior findings on the drivers of citations are plentiful, they are often inconsistent between samples and disciplines. For example, the aforementioned link between the length of the abstract and citations appears to exist in medicine but not in sociology~\cite{vanWesel.2014}. The association between the length of the title and citations appears to be positive in the field of medicine~\cite{vanWesel.2014}, absent in biology and related disciplines~\cite{Jamali.2011} as well as addiction research~\cite{Rostami.2014}, and even negative in applied physics~\cite{vanWesel.2014} and marketing~\cite{Stremersch.2015}.

In this article, we contribute by proposing a parsimonious theoretical two-step filtering process model of how citations come to be with respect to article metadata, and by systematically assessing corresponding metadata drivers of citations in articles from some key venues in the field of software engineering. We identify relevant antecedents regarding articles' text (titles, abstracts, and keywords) and author teams.

In the following, we first provide background on related research, the we develop explicit hypotheses, describe our study design, report our empirical results, discuss them, and conclude with an outlook on future research.

\section{Background and related work}

Our central research question is: \textit{Which readily observable characteristics of a software engineering article's metadata (title, abstract, keywords, and author team) drive citations to the focal article?}

There are at least three streams of research that relate to this question. First, there is extensive research on the drivers of citations in fields outside of software engineering. In fact, researchers in different disciplines have identified at least 28 different antecedents of citation count~\cite{Tahamtan.2016, Onodera.2015, Onodera.2021}. Specifically, there appear to be three broad categories of factors that influence citations: paper-related, journal-related, and author-related~\cite{Tahamtan.2016}.

On paper-related factors, authors have, for instance, studied characteristics of titles, abstracts,  keywords, and the manuscript text of articles. 
\citeauthor{Ayres.2000}~\cite{Ayres.2000} found that manuscript length has a positive effect on citations in legal scholarship, whereas title length has a negative effect.
\citeauthor{Stremersch.2015}~\cite{Stremersch.2015} found a similar negative effect of title length in marketing, as did \citeauthor{Subotic.2014}~\cite{Subotic.2014} in psychology.
Van Wesel et al.~\cite{vanWesel.2014} showed that abstract length positively affects citations in medicine and applied physics, but not in sociology. 
\citeauthor{Buter.2011}~\cite{Buter.2011} demonstrated that punctuation marks such as hyphens and colons in titles have a small positive effect on citations in a wide range of disciplines, including clinical medicine and food science, but a negative effect in others like biology.
\citeauthor{Jamali.2011}~\cite{Jamali.2011} presented evidence that articles with question marks in the title are downloaded more frequently but ultimately less cited in the field of astronomy.
\citeauthor{Chakraborty.2014}~\cite{Chakraborty.2014} and \citeauthor{Rostami.2014}~\cite{Rostami.2014} showed that the diversity of keywords positively predicts citations in a large multidisciplinary sample of articles. Relatedly, \citeauthor{So.2015}~\cite{So.2015} found, among other things, that the number of keywords is positively associated with citations.
Regarding the manuscript text of an article, \citeauthor{vanWesel.2014}~\cite{vanWesel.2014} identified the number of pages, the number of sentences, and the number of cited references as factors influencing citations, but they observed substantial variance across academic disciplines.

Regarding journal-related factors, \citeauthor{vanDalen.2005}~\cite{vanDalen.2005} found that the reputation of a journal (e.g., the journal impact factor or citations of editorial board members) is positively linked to citations of articles appearing in the focal journal in demography. Many others found similar results in other fields (e.g.,~\cite{Lariviere.2010, Wang.2011}).

Several author-related factors have also received attention. For instance, the number of authors has frequently been studied. Van Wesel et al.~\cite{vanWesel.2014} found a positive relationship with citations in sociology, medicine, and applied physics. \citeauthor{Ibanez.2013}~\cite{Ibanez.2013} studied Spanish computer science professors and found that papers by two authors receive the most citations, and that international collaboration is also positively associated with citations. \citeauthor{Stremersch.2007}~\cite{Stremersch.2007} found at least partly contrasting evidence in marketing, suggesting that an increasing number of authors is linked to fewer citations. \citeauthor{Stremersch.2007}~\cite{Stremersch.2007} also showed evidence that U.S. affiliations of authors do not drive citations, whereas their own prior work~\cite{Stremersch.2005} and that of others suggested~\cite{vanDalen.2005} such a link. Another oft-studied characteristic is the reputation of authors. Van Dalen and Henkens~\cite{vanDalen.2005}, for instance, found that author reputation in the form of accumulated past citations is positively related to citations.

It is worth noting that there exists extensive literature across disciplines and that---as the selected examples above already indicate---there is great variance in the results. For almost every significant finding, there are studies where the same characteristic is not significant or even significant in the opposite direction~\cite{Tahamtan.2016}.

Second, there is a rich bibliometric literature in the field of software engineering. For one, researchers have shown an interest in identifying most-cited papers. For example, \citeauthor{Wohlin.2005}~\cite{Wohlin.2005, Wohlin.2007, Wohlin.2008} conducted a well-known series of studies that identified the most-cited papers in software engineering. \citeauthor{Kitchenham.2010}~\cite{Kitchenham.2010} also analyzed the key topics, objectives, and content of the papers. \citeauthor{Garousi.2016b}~\cite{Garousi.2016b} identified the top 100 most-cited papers in software engineering. For another, several bibliometric studies attempted to identify not top papers but top scholars and institutions in the field of software engineering in various geographies. \citeauthor{Garousi.2010}~\cite{Garousi.2010}, for instance, performed such an analysis for Canada, whereas \citeauthor{Hummel.2012}~\cite{Hummel.2012} took a global approach and ranked highly influential software engineering scholars who were program committee or organizing committee members of reputable software engineering venues or ACM SIGSOFT research award winners. Other authors fused these two substreams of research, reporting on both top papers and scholars~\cite{Chuang.2014}. Yet another substream takes a broader yet still primarily descriptive perspective on citations. \citeauthor{Garousi.2016}~\cite{Garousi.2016}, for instance, performed a host of descriptive analyses of the citation distributions in software engineering, the studied topics in articles, as well as the field's publication output over time, while also presenting evidence that review articles are more frequently cited than other article types and that paper age is related to citations. \citeauthor{Ahmad.2020}~\cite{Ahmad.2020, Ahmad.2019} studied \textit{XP Conference} contributions and identify most-cited papers and recently 
``hot'' topics, mapped out citation distributions, and showed linkages between the number of authors and article age and the number of citations.

Finally, the last stream of related research is the intersection of the aforementioned streams, and thus directly concerns itself with the antecedents of citations in software engineering.\footnote{Note that some of the articles listed in the category above also occasionally link article characteristics to citations (e.g.,~\cite{Kitchenham.2010, Ahmad.2020, Garousi.2016}). We nevertheless classified them as bibliometric studies because their primary interest was  not the identification of drivers of citations.} Specifically, this nascent stream contains several papers of note, which, however, all have relevant limitations.

\citeauthor{Garousi.2017}~\cite{Garousi.2017} studied how different drivers impact citation frequency in software engineering. Specifically, they analyzed the citation impact of different types of venues (journals vs. conferences), different venues, authors' countries of origin, and ``top-10 authors'' (as measured by authors' total number of papers). While this study is highly instructive, it is exploratory as it treated each driver in isolation and did not combine all drivers into one model to account for potentially confounding effects between them.

\citeauthor{Molleri.2018}~\cite{Molleri.2018} analyzed the relationship between articles' reporting quality regarding relevance and rigor, and citations. They found no effect of relevance, but could demonstrate an association between reported rigor and citations. However, they did so without controlling for the potential influence of other drivers like article length, which they found to be positively correlated with rigor.

The paper most closely related to our work is a recent working paper by \citeauthor{Mantyla.2019}~\cite{Mantyla.2019}. They found that venues, past citations of authors, article length, and the number and recency of references in an article particularly affect the citations an article attracts. While this study doubtlessly substantially advances the field, it has not been formally reviewed yet, and it considered only journal publications, neglecting conference papers.

In addition, none of these prior studies engaged in formal theory testing~\cite{Popper.2005}. While they all provide well-argued reasons for why certain factors might affect citations, none of them truly specified models ex ante and tested them. In fact, the majority of these articles did not jointly test a comprehensive model of drivers, but instead tested each variable in isolation, potentially missing confounding effects between multiple variables. \citeauthor{Mantyla.2019}~\cite{Mantyla.2019} overcame this limitation but introduced new variables into their regression models in a step-wise fashion. While this is helpful and appropriate for creating a robust set of predictors of citations, it is epistemologically at odds with our objective of building and testing a theoretical model. 

Our work differs from prior work in the following ways. First, we attempt to develop a simple theoretical model of citations, explicitly theorize on drivers of citations, and then subject our hypotheses to empirical testing. Second, in this quest, we purposefully focus on the ``early stage'' of the citation process. As we detail below, we are particularly interested in metadata that is apparent from search results in scientific search engines, before a potentially citing researcher has read the entire paper. In other words, we are interested in a subset of what \citeauthor{Onodera.2015} label ``extrinsic factors''~\cite{Onodera.2015, Onodera.2021}. Third, we employ multiple regression analyses to account for multiple article characteristics at the same time, attempting to control for a wide range of potentially confounding factors. Finally, we introduce a hitherto unstudied characteristic of abstracts to the study of citations, namely abstracts' propositional idea density.

\section{Hypotheses}

Out of the many possible metadata characteristics of articles, we selected several for inclusion in this study. Our guiding logic here was (a) that these characteristics should be within the realm of influence of the article authors, (b) that we suspected a potential causal link to citation counts, rather than mere correlations, and (c) that the necessary data be available for a large-scale analysis. Specifically, we opted for characteristics related to the author team and the article itself, which are two of the three categories identified by \citeauthor{Tahamtan.2016}~\cite{Tahamtan.2016}. We decided to exclude the third category, i.e., journal characteristics such as a journal's impact factor. The reason for this is not that we believe these characteristics have no influence. To the contrary, we believe that there are likely so many strong effects that we should capture them comprehensively through fixed effects, as we will explain in the method section. This does not allow us to hypothesize on and meaningfully test such effects individually, but it enables us to adequately control any confounding effects that might originate from them, thus strengthening our hypothesis tests regarding the other categories.

\begin{figure}[h]
  \centering
  \includegraphics[width=\linewidth]{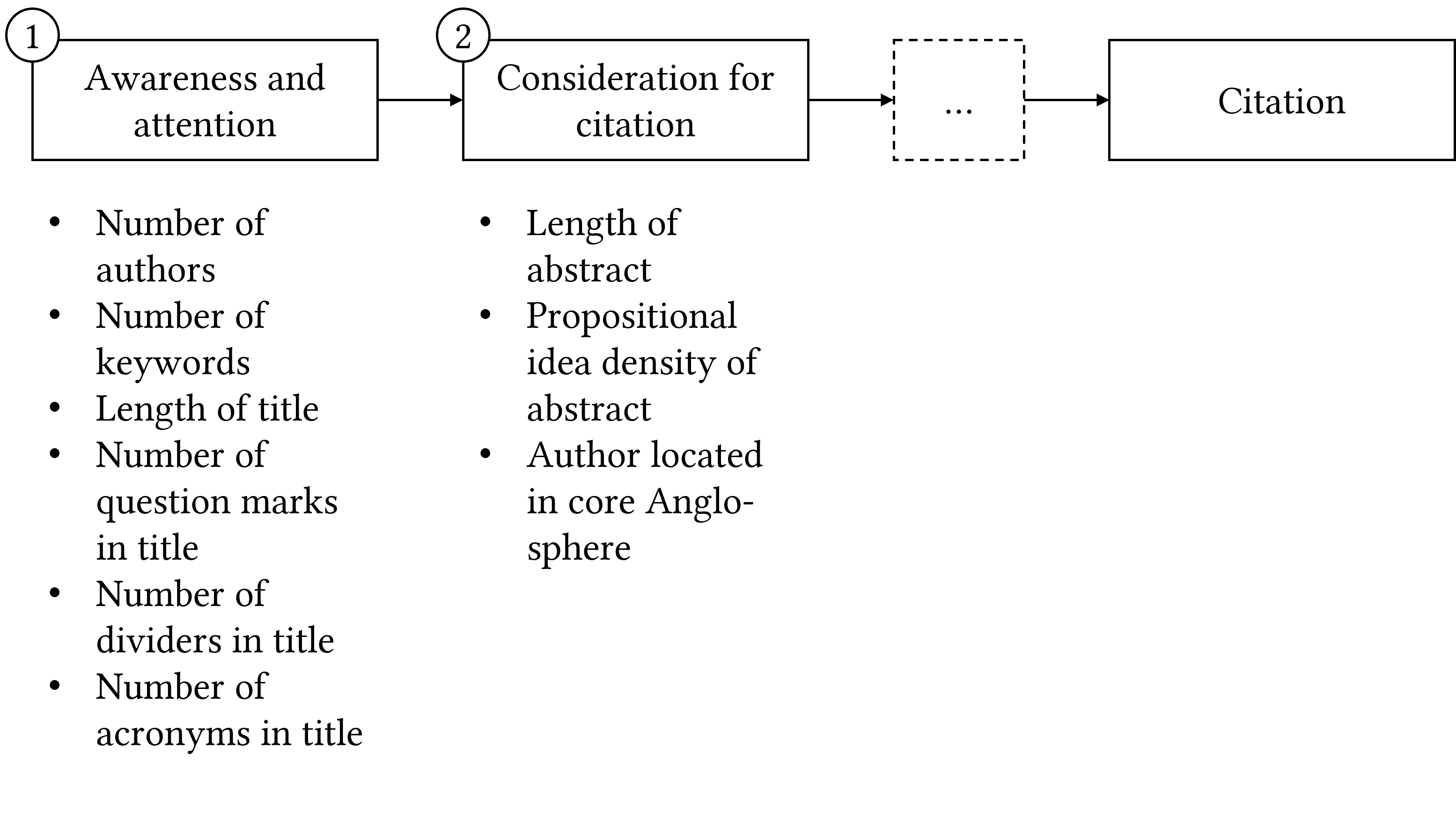}
  \caption{Filtering process\label{fig1}}
\end{figure}

We propose that there is a two-step filtering process that needs to be passed for an article to be cited (see Figure \ref{fig1}). In a first step, readers must become aware of and attentive to an article to at least consider it citeable. Particularly easily accessible metadata that may, for instance, be shown as results by scientific search engines or that may influence the inclusion of an article in search results---like the article's author(s), keywords, and title---are likely to play a role here. In the second step, readers must be compelled to cite the focal article in their own work. In this step, readers are likely to consult at least the abstract of the article. The greater the likelihood that any given article passes each filter, the greater the likelihood that the article gets cited, and thus the greater its total citations. Note that we purposefully focus on these early steps of the citation process, which leads us to not study other potentially relevant variables that might affect citations but that are not influential in the early stages, e.g., the number of pages in an article~\cite{Mantyla.2019} or the quality of an article's content~\cite{Molleri.2018}. In the following, we develop explicit hypotheses about several characteristics that may influence these two steps.

In the first step of the filtering process, awareness and attention, we propose that the \textit{number of authors} has a positive influence. If the author team is larger, this might not only lead to higher-quality work due to the inclusion of diverse perspectives~\cite{Padial.2010}, or to more self-citations~\cite{Herbertz.1995}, but a larger author team will also have more opportunities through which to promote their article~\cite{Stremersch.2007, Leimu.2005}. Specifically, more authors can increase 
``knowledge diffusion''~\cite[p. 832]{Bosquet.2013} through presenting their work at more different conferences or workshops, and promoting it more through informal, personal channels~\cite{Bosquet.2013}. In line with much of prior research, e.g.,~\cite{Aksnes.2003, vanWesel.2014, So.2015}, we thus propose:

\textbf{Hypothesis 1a (H1a):} \textit{There is a positive association between the number of authors and citations.}

Another way in which authors can make their work visible to potential readers is by including a \textit{number of keywords}. When an article includes many keywords, the article has a greater chance of showing up in searches~\cite{Rostami.2014}, and it might thus have a greater chance of passing the first filtering step. In line with prior research~\cite{So.2015}, we therefore propose:

\textbf{Hypothesis 1b (H1b):} \textit{There is a positive association between the number of keywords and citations.}

We next hypothesize a link between the \textit{length of an article's title} and citations. Specifically, we argue that longer titles have the potential to convey more, and more specific, information, making the article potentially more attractive to readers. Consequently, we hypothesize in line with some prior results~\cite{Rostami.2014, Jacques.2010}:

\textbf{Hypothesis 1c (H1c):} \textit{There is a positive association between the length of the title and citations.}

Another way in which authors might make their papers more appealing is by using certain rhetorical devices. Specifically, authors frequently pose questions (that they may or may not answer in the title), and thus use \textit{question marks}, in an article's title. Research on citations in computer science overall~\cite{Fiala.2021} and news media~\cite{Lai.2014} has shown that the use of questions increases engagement with an article (``the clickbait phenomenon''~\cite[p. 1301]{Kuiken.2017}). We propose that the presence of such questions make an article more interesting and thus more likely to be read and considered by readers. Put formally:

\textbf{Hypothesis 1d (H1d):} \textit{There is a positive association between the number of question marks in the title and citations.}

Similarly, authors often use \textit{dividers} such as a colon or a dash in article titles. Composing such ``combinational titles''~\cite[p. 2007]{Rostami.2014} allows them to tease an interesting story (e.g., ``Understanding IV\&V in a safety critical and complex evolutionary environment: The NASA Space Shuttle program''), communicate the results in a vivid way (e.g., ``Same App, Different Countries: A Preliminary User Reviews Study on Most Downloaded iOS Apps'' or ``Exception Analysis and Points-to Analysis: Better Together''), introduce a potentially useful named artifact (e.g., ``MMLT: A mutual multilevel trust framework based on trusted third parties in multicloud environments''), or include methodological details like the type of study (e.g., ``Software-testing education: A systematic literature mapping'' or ``How does combinatorial testing perform in the real world: an empirical study''). In addition, a colon or a dash allows the authors to provide more information, given that they can effectively include two sentences in the title instead of one. All of these cases are likely to make a paper more appealing and thus pass the first filtering process. Compatible with prior research~\cite{Rostami.2014, Jacques.2010}, we thus propose:

\textbf{Hypothesis 1e (H1e):} \textit{There is a positive association between the number of dividers in the title and citations.}

In contrast to the prior characteristics, we propose that a greater \textit{number of acronyms} in a title will make a paper, in general, less relevant and appealing to readers. Notably, this goes against prior findings in the medical field. However, these findings are likely due to particularities of that field, where acronyms may characterize articles from highly influential multi-center trials or organizations, or about specific genes~\cite{Jacques.2010}. We instead argue that, for one, acronyms might be hard (or impossible) to comprehend. For another, acronyms might signal a more specialized or narrower topic that likely only appeals to a smaller subset of readers. We thus formalize:

\textbf{Hypothesis 1f (H1f):} \textit{There is a negative association between the number of acronyms in the title and citations.}

We now turn to the second filtering step of Figure~\ref{fig1}, consideration for citation. In this step, readers have chosen to engage with an article at least far enough to read the abstract, and might be able to make a decision whether to cite it in their own work. One key determinant is likely the information content of the abstract. Specifically, the greater the \textit{length of an abstract}, the more information it can contain and the more it might give the reader a good sense for an article, potentially even allowing the reader to cite the article without reading the entire text~\cite{vanWesel.2014, Janes.1991}. Consequently, we argue:

\textbf{Hypothesis 2a (H2a):} \textit{There is a positive association between the length of the abstract and citations.}

Another factor we consider is the \textit{propositional idea density~\cite{Covington.2009} of the abstract}. Propositional idea density captures the number of propositions or ideas, normalized by the number of words in a text. From the perspective of a stream of psycholinguistic research going back to seminal research by Kintsch and colleagues~\cite{Kintsch.1974, Kintsch.1973}, an idea is whatever can be true or false, and thus the basic unit involved in the understanding and retention of text. Consequently, higher idea density makes text harder to understand, but also richer in content (when keeping text length constant)~\cite{Kintsch.1974, Kintsch.1973, Kintsch.1998}. For example, the two sentences ``the previously obsolete system is relevant again'' and ``there is a system which was previously obsolete, and it is relevant again'' have the same informational content but differ in their idea density. Prior literature has linked propositional idea density to cognitive functioning and impairment, even demonstrating, for example, its suitability as a screening tool for detecting Alzheimer's disease based on a person's writing~\cite{Riley.2005, Hill.2021}. Conversely, we suggest that greater idea density in an abstract indicates a greater density of content and thus a larger number of reasons for which an article could be cited (independent of abstract length), thus leading to a higher number of citations~\cite{Runco.2017, Forthmann.2020}. We posit:

\textbf{Hypothesis 2b (H2b):} \textit{There is a positive association between the propositional idea density of the abstract and citations.}

Finally, there might be an additional effect of the authors on citations, which was found in prior work outside software engineering with regard to U.S.-based authors~\cite{Stremersch.2005} and the Anglo-Saxon world~\cite{Peng.2012}. Specifically, the entire field of computer science is substantially influenced by institutions and scholars that are based in the Anglo-Saxon world. Citing papers with \textit{corresponding authors based in the ``core Anglosphere''}~\cite{Wellings.2019}, i.e., the developed Anglo-Saxon countries, might provide legitimacy to the citing work, as it will signal that the articles build on ideas from a critical country in the field. Authors at such institutions are also more likely to be native English speakers, potentially leading to more elegantly written abstracts and even articles, which again might increase citation rates. Put formally:

\textbf{Hypothesis 2c (H2c):} \textit{There is a positive association between the corresponding author being located in the core Anglosphere and citations.}

\section{Study design}

In the following section, we detail the study design. Specifically, we discuss our sample and data, explain how we operationalized our variables, and describe the data analysis.

\subsection{Sample and data}

We obtained a sample of article abstracts and titles from Clarivate's Web of Science (WoS, formerly known as Web of Knowledge), a database that has been used in recent studies analyzing textual characteristics and/or citations of articles in computer science, information systems, and other disciplines~\cite{vanWesel.2014, GrafVlachy.2021, Fiala.2021, Buter.2011, Onodera.2021, Onodera.2015, GrafVlachy.2021b}. We opted for WoS since it is superior to Google Scholar in various ways~\cite{Tahamtan.2016, Garousi.2016} and we did not have access to Scopus. We manually downloaded all WoS records of articles published between 1990 and 2020 and indexed in the category ``Computer Science - Software Engineering'' as of May 2021. Since this category also includes a host of venues that are not primarily or even  tangentially related to software engineering, such as, for example, the \textit{USENIX Workshop on Steps to Reducing Unwanted Traffic on the Internet}, we sought to appropriately restrict our sample. Specifically, we identified the top 20 venues according to Google Scholar from the category ``Engineering \& Computer Science - Software Systems'' as of July 2021 (see Table \ref{tabVenues}), and only retained articles from these venues.

\begin{table*}[ht]
\small
\caption{Software engineering venues included in analysis \label{tabVenues}}
\begin{tabular}{clc}

\toprule
\# & Publication & h5-index \\
\midrule
1 & ACM/IEEE International Conference on Software Engineering & 74 \\
2 & Journal of Systems and Software & 61 \\
3 & Information and Software Technology & 59 \\
4 & ACM SIGSOFT International Symposium on Foundations of Software Engineering & 53 \\
5 & Empirical Software Engineering & 53 \\
6 & IEEE Transactions on Software Engineering & 52 \\
7 & ACM SIGPLAN-SIGACT Symposium on Principles of Programming Languages (POPL) & 48 \\
8 & ACM SIGPLAN Conference on Programming Language Design and Implementation (PLDI) & 46 \\
9 & IEEE/ACM International Conference on Automated Software Engineering (ASE) & 45 \\
10 & IEEE Software & 44 \\
11 & Symposium on Operating Systems Principles & 42 \\
12 & Software \& Systems Modeling & 41 \\
13 & Mining Software Repositories & 40 \\
14 & International Conference on Software Analysis, Evolution, and Reengineering (SANER) & 40 \\
15 & International Symposium on Software Testing and Analysis & 36 \\
16 & International Conference on Tools and Algorithms for the Construction and Analysis of Systems (TACAS) & 33 \\
17 & IEEE International Conference on Software Maintenance and Evolution & 33 \\
18 & Proceedings of the ACM on Programming Languages & 31 \\
19 & Software: Practice and Experience & 30 \\
20 & ACM SIGPLAN International Conference on Object-Oriented Programming, Systems, Languages, and Applications (OOPSLA) & 29 \\
\bottomrule
\end{tabular}
\end{table*}

We restricted our sample to articles published after 1990 because the coverage of WoS is very limited before this time~\cite{GrafVlachy.2021}. We further restricted our sample to articles published until 2020 for two reasons. First, at the time of data collection, the following year was not finished, and we would thus have necessarily obtained an incomplete sample for that year. Second, since our variable of interest is the number of citations, we can only analyze articles that had a realistic chance of getting cited in the first place. The time window for citations would have been extremely narrow for the most recent articles and would very likely only have introduced more noise into our sample than contribute explanatory value.

We dropped duplicate entries from WoS, removed all records that did not pertain to an ``Article,'' ``Article; Proceedings Paper,'' or ``Proceedings Paper,'' thereby removing other types of publications like book reviews or editorials.

As the same venue might be included in WoS under a host of different names (e.g., \textit{OOPSLA 11: Proceedings of the 2011 ACM International Conference on Object Oriented Programming Systems Languages and Applications} and \textit{OOPSLA 2008 Nashville, Conference Proceedings}) we performed data cleaning and manually identified all records that belong to the 20 venues.

To further clean the data, we performed minor preprocessing on the abstracts (but not the titles) using Perl scripts.\footnote{All used scripts are available at \url{https://doi.org/10.6084/m9.figshare.17209178}.} Specifically, we removed copyright and classification strings that were frequently included, we expanded some very common non-technical abbreviations like ``e.g.'' removed enumerations, and made adjustments to facilitate later analysis (such as converting colons to full stops and removing full stops in ``et al.'' to aid sentence recognition). We manually inspected 500 randomly sampled abstracts to ensure that our preprocessing did not have unintended side effects.

We dropped all abstracts with less than three full stops in them because spot checks indicated that these cases were errors in WoS, e.g., abstracts which contained no punctuation at all. We also dropped all articles which had titles with a length of  three or less characters as these were clear errors in WoS.

Our sample in the 20 selected venues comprised 22,984 articles. Of these, we had sufficient data for 16,131 to use in our analyses.

\subsection{Operationalization of variables}

We measure our dependent variable, the \textit{number of citations}, simply as the total number of citations an article had accrued at the time of data collection, as recorded by WoS. Unlike some prior researchers, we do not log-transform the number of citations~\cite{vanWesel.2014}. This is because this variable is a count, which has distributional properties that are best addressed by using an appropriate estimator rather than performing transformations~\cite{Gardner.1995}. Specifically, transformations cannot produce normally distributed errors which would be necessary for, for instance, ordinary least squares regression.

We measured our independent variables as follows. The \textit{number of authors} is the count of authors as listed in WoS. The \textit{number of keywords} is measured the same way. The \textit{length of the title} is the number of characters in the article's title, divided by 100 to make regression coefficients more interpretable. The \textit{number of question marks in the title} is again a simple count. The \textit{number of dividers in the title} is the count of colons followed by a blank space plus the count of dashes or double dashes preceded and succeeded by a blank space (to avoid incorrectly counting hyphenated words as in ``3D-Color-Structure-Code - A new non-plainness island hierarchy'' or ``A Computation- and Communication- Infrastructure for Modular Special Instructions in a Dynamically Reconfigurable Processor''). The \textit{number of acronyms in the title} was determined as the total number of all-caps words consisting of at least two characters (with or without separating full stops) in a title, while setting this value to zero for articles that had titles that were entirely in all-caps. The \textit{length of the abstract} was the number of characters in the abstract before any preprocessing, again divided by 100 for interpretability.

To measure the \textit{propositional idea density of the abstract}, we use CPIDR (Computerized Propositional Idea Density Rater) 5, a software specifically built and validated for this purpose~\cite{Brown.2008}, as well as used in prior research~\cite{Ferguson.2014, Hill.2021}. 
The software exploits the fact that propositional idea density---as defined in the present article~\cite{Kintsch.1974, Turner.1977}---can be approximated by relying on syntactic features of text. Specifically, performing part-of-speech tagging of text can provide approximations of the frequencies of various categories of words. Summing up the number of verbs, adjectives, adverbs, prepositions, and conjunctions, and dividing the sum by the total number of words then yields the propositional idea density of a text~\cite{Snowdon.1996}. Note that Kintsch's notion of propositional idea density~\cite{Kintsch.1974} differs from formal semantics in that common nouns are not considered propositions. Similarly, information about verb tense, modality, or co-reference does not enter the count of propositions.
The developers of the CPIDR software validated it in two ways~\cite{Brown.2008}. First, they established that the software essentially replicates some examples provided in the literature on propositional idea density~\cite{Turner.1977}. Second, they tested the software against human raters and found that the results of the software correlated very highly (\textit{r} = .97) with consensus ratings of human raters~\cite{Brown.2008}.
We standardized the value of propositional idea density for easier interpretation.

To determine whether the \textit{corresponding author was located in the core Anglosphere}, we identified the institutional affiliation of the author listed as the ``reprint author'' (which indicates the corresponding author) in WoS, and created a binary variable set to one if this institution was located in the U.S., Canada, the U.K., Ireland, Australia, or New Zealand~\cite{Wellings.2019}, and to zero otherwise.\footnote{Note that because our argument is largely based on the specific legitimacy-granting power of these countries, we deliberately do not consider other countries that have English as a \textit{de jure} or \textit{de facto} official language.}

\subsection{Data analysis}

To account for the non-independence of observations at the venue-level, we select a fixed-effects Poisson estimator and calculate standard errors clustered at the venue. This estimator is appropriate for analyses with a dependent variable that is a count~\cite{Gardner.1995} such as the number of a paper's citations. In addition, this estimator is robust to a host of potential problems and specifically more robust than a negative binomial estimator~\cite{Wooldridge.1999, Wooldridge.2010}. Our choice to use cluster-robust standard errors accounts for the non-independence of observations, i.e., the fact that that error terms in the regression model might be correlated for articles because they were published in the same venue.

We also include an extensive array of so-called \textit{dummy variables} in our models. Dummy variables are binary variables that are included in regression models to represent categorical values~\cite{Hardy.1993}. Typically, dummy variables are used to account for so-called ``fixed effects'' of specific types of observations~\cite{Wooldridge.2010}, e.g., the effects of the specific venue in which an article was published~\cite{Mantyla.2019}. To this end, a dummy variable is created for every venue in the data set, and for every article only the corresponding dummy variable is set to one, with all others taking the value zero.\footnote{To avoid perfect multicollinearity in the regression, one of the dummy variables must be deliberately omitted. This is done automatically by the statistical software we use.}

We include dummies for three reasons. First, we decided to not normalize citation counts by the age of a paper (e.g.,~\cite{Molleri.2018, Garousi.2016}) because this requires using a ratio as a dependent variable, which is not advisable in regression models~\cite{Certo.2020}. Instead, we can include dummy variables for every year, which explicitly accounts for an article's age, which influences citations for obvious reasons~\cite{Tahamtan.2016, Mantyla.2019, Ahmad.2020, Garousi.2016}. This solution has the added advantage that it accounts for all other effects that pertain to a specific year as well, e.g., the volume of total scientific output or total citations made by all researchers in a given year, and it does not assume that any time effects be linear.

Second, as journal articles are typically cited more frequently than conference papers~\cite{Garousi.2016}, and there are systematic differences between different subfields of a discipline~\cite{Tahamtan.2016} as well as venues~\cite{Ibanez.2013, Tahamtan.2016, Mantyla.2019}, it would further seem prudent to include dummy variables for each venue~\cite{vanDalen.2005}. This way, they control for all time-invariant effects of all venues.

Finally, there might be effects that are fixed for every venue-year combination. We thus included interactions (i.e., multiplications) between all year and venue dummy variables. This allows us to account for all effects that might be related to all articles appearing in a venue in a specific year, capturing, e.g., the effects of changing journal or conference prestige or impact factors over time. Note that our model thus explicitly accounts for all these variables, which have often been studied in prior literature~\cite{Mantyla.2019}. We deem their importance so uncontroversial that we did not hypothesize about them. Instead, we incorporate them in our model in a way that not only controls for their specific effects, but that also allows us to control for all other effects that are due to particularities of venues in specific years (and which prior literature typically does not account for). This includes, for instance, editorial policies or regulations regarding maximum abstract or title lengths. Our model is thus, in this way, more comprehensive than prior work in software engineering.\footnote{Because these dummy variables contain all the information contained in the aforementioned individual dummy variables for venues only and for years only, these individual variables are automatically omitted from our regression analyses due to perfect multicollinearity without any loss of information.}

We performed all statistical analyses using Stata 17.0.

\section{Results}

Table \ref{desc1} displays key descriptive statistics for the main variables. Correlations between them are shown in Table \ref{corr2}. Variance inflation factors were all far below commonly used thresholds, alleviating concerns over multicollinearity~\cite{Obrien.2007}.

Table \ref{reg1} shows the regression results.\footnote{We do not report the coefficients of the dummy variables as their interpretation is likely not helpful in our setting. Given the inclusion of venue-year dummies, we could study questions like ``Is the effect of journal A in year X greater than the effect of conference B in year Y?,'' but we cannot, for example, make blanket comparisons between venues. A manual inspection of the dummy variables shows that many of them are significant, confirming our initial assumption that there are meaningful effects of publication venue and year that should be accounted for.}
We find a positive association between the number of authors and citations, supporting H1a. The number of keywords is only marginally related to citations, and negatively at that, providing weak evidence against H1b. We also find no support for H1c, which suggested a link between the length of an article's title and citations. Citations were, however, significantly positively linked to the number of questions marks (supporting H1d) and dividers (supporting H1e) in abstract titles. The hypothesized negative effect of acronyms in the title on citations is also borne out in our model, supporting H1f. Regarding articles' abstracts, we find support for H2a and H2b, since both length and propositional idea density of the abstract were significantly positively related to citations. Finally, an affiliation of the corresponding author with an institution in the core Anglosphere is also linked to citations, supporting H2c.

Our results are robust to alternative model specifications. For instance, all coefficients that are significant in our primary analysis are also significant in a negative binomial regression with robust standard errors. In addition, the coefficient for the number of keywords, which was insignificant in the main analysis, also becomes significant in a negative binomial model.

Although concerns over multiple hypothesis testing are not common in the literature on citations, we performed an additional robustness check to control the false discovery rate and computed adjusted \textit{p}-values (sharpened \textit{q}-values)~\cite{Benjamini.2006, Anderson.2008} for all variables of interest. The results essentially reaffirm our original findings, even strengthening the result for the number of keywords, and letting the coefficient of title length reach marginal significance.

\begin{table}[htbp]\centering \caption{Descriptive statistics of all variables \label{desc1}} %
\def\sym#1{\ifmmode^{#1}\else\(^{#1}\)\fi}
\small %
\begin{tabular}{lcccccc}
\toprule

                    &        Mean&          SD&         Min.&         Max.\\
\midrule
Citations           &       17.63&       46.64&        0.00&    2,301.00\\
Authors (log)       &        1.08&        0.52&        0.00&        3.50\\
Keywords            &        4.49&        1.68&        1.00&       23.00\\
Length (title)      &        0.72&        0.24&        0.01&        2.01\\
Question marks (title)&        0.03&        0.17&        0.00&        3.00\\
Dividers (title)    &        0.29&        0.46&        0.00&        2.00\\
Acronyms (title)    &        0.23&        0.51&        0.00&        6.00\\
Length (abstract)   &       12.18&        4.18&        2.43&       41.37\\
Idea density (abstract)&        0.00&        1.00&       -3.93&        4.85\\
Anglosphere author  &        0.39&        0.49&        0.00&        1.00\\

\bottomrule
\multicolumn{2}{l}{\footnotesize SD = Standard deviation}\\
\multicolumn{2}{l}{\footnotesize \textit{N} = 16,131}\\
\end{tabular}
\end{table}

\begin{table*}[ht]
\def\sym#1{\ifmmode^{#1}\else\(^{#1}\)\fi}
\small
\caption{Correlations between all variables \label{corr2}}
\begin{tabular}
{lcccccccccc}
\toprule

                &Citations         &\makecell{Authors\\(log)}         & \makecell{Key-\\words}         &\makecell{Length\\(title)}         &\makecell{Question\\marks\\(title)}         &\makecell{Dividers\\(title)}         &\makecell{Acronyms\\(title)}         &\makecell{Length\\(abstract)}         &\makecell{Idea\\density\\(abstract)}         &\makecell{Anglo-\\sphere\\author}         \\
\hline
Citations       &     1.00         &                  &                  &                  &                  &                  &                  &                  &                  &                  \\
Authors (log)   &     0.02         &     1.00         &                  &                  &                  &                  &                  &                  &                  &                  \\
Keywords        &     0.03\sym{***}&    -0.03\sym{***}&     1.00         &                  &                  &                  &                  &                  &                  &                  \\
Length (title)  &    -0.02\sym{*}  &     0.10\sym{***}&     0.09\sym{***}&     1.00         &                  &                  &                  &                  &                  &                  \\
Question marks (title)&     0.01         &     0.04\sym{***}&    -0.01         &     0.05\sym{***}&     1.00         &                  &                  &                  &                  &                  \\
Dividers (title)&    -0.01         &     0.09\sym{***}&    -0.01         &     0.22\sym{***}&    -0.02\sym{*}  &     1.00         &                  &                  &                  &                  \\
Acronyms (title)&    -0.04\sym{***}&     0.05\sym{***}&     0.02\sym{*}  &     0.03\sym{**} &    -0.04\sym{***}&     0.18\sym{***}&     1.00         &                  &                  &                  \\
Length (abstract)&     0.00         &     0.18\sym{***}&     0.11\sym{***}&     0.20\sym{***}&     0.05\sym{***}&     0.04\sym{***}&    -0.02\sym{**} &     1.00         &                  &                  \\
Idea density (abstract)&     0.01         &     0.02\sym{**} &    -0.03\sym{***}&    -0.07\sym{***}&     0.06\sym{***}&     0.02\sym{*}  &    -0.11\sym{***}&     0.03\sym{***}&     1.00         &                  \\
Anglosphere author&     0.07\sym{***}&    -0.08\sym{***}&     0.01         &    -0.11\sym{***}&     0.01         &     0.00         &    -0.05\sym{***}&    -0.07\sym{***}&     0.07\sym{***}&     1.00         \\

\bottomrule
\multicolumn{11}{l}{\footnotesize \sym{*} \(p<0.05\), \sym{**} \(p<0.01\), \sym{***} \(p<0.001\)}\\
\end{tabular}
\end{table*}

\begin{table}[ht]
\def\sym#1{\ifmmode^{#1}\else\(^{#1}\)\fi}
\small
\caption{Drivers of citations \label{reg1}}
\begin{tabular}{lcc}
\toprule

                    & Coefficient & IRR\\
\midrule
Authors (log)       &       0.331\sym{***}    & 1.392\\
                    &     (0.047)         &\\
\addlinespace
Keywords            &      -0.016\sym{+}      & 0.984\\
                    &     (0.009)         &\\
\addlinespace
Length (title)      &      -0.217             & 0.805\\
                    &     (0.190)         &\\
\addlinespace
Question marks (title)&       0.414\sym{***}    & 1.512\\
                    &     (0.065)         &\\
\addlinespace
Dividers (title)    &       0.150\sym{*}      & 1.161\\
                    &     (0.063)         &\\
\addlinespace
Acronyms (title)    &      -0.192\sym{***}    & 0.826\\
                    &     (0.058)         &\\
\addlinespace
Length (abstract)   &       0.017\sym{*}      & 1.017\\
                    &     (0.008)         &\\
\addlinespace
Idea density (abstract)&       0.068\sym{***}    & 1.070\\
                    &     (0.016)         &\\
\addlinespace
Anglosphere author  &       0.161\sym{***}    & 1.174\\
                    &     (0.038)         &\\
\midrule
Observations        &         \multicolumn{2}{c}{16,131}\\
Venue fixed effects            &         \multicolumn{2}{c}{Yes}\\
Year fixed effects             &         \multicolumn{2}{c}{Yes}\\
Venue-year fixed effects       &         \multicolumn{2}{c}{Yes}\\

\bottomrule
\multicolumn{3}{l}{\footnotesize Fixed-effects Poisson regression}\\
\multicolumn{3}{l}{\footnotesize Cluster-robust standard errors (by venue) in parentheses}\\
\multicolumn{3}{l}{\footnotesize IRR = Incidence rate ratio}\\
\multicolumn{3}{l}{\footnotesize \sym{+} \(p<0.10\), \sym{*} \(p<0.05\), \sym{**} \(p<0.01\), \sym{***} \(p<0.001\)}\\
\end{tabular}
\end{table}

\section{Discussion}

In the following, we discuss implications as well as threats to internal and external validity.

\subsection{Implications}

Our empirical results have several important implications.

\textbf{Implication \#1:} Early-stage metadata characteristics drive citations in software engineering. Our results show that citations can be at least partially explained by metadata characteristics that might function as ``signals''~\cite{vanDalen.2005} and that likely have a particularly meaningful effect at the early stages of the literature review process.

\textbf{Implication \#2:} Propositional idea density matters. As the first study in software engineering, we identified a link between the idea density of articles' abstracts and the citations they receive.

\textbf{Implication \#3:} The software engineering discipline behaves differently. Despite many findings confirming prior results in other disciplines, there is a marked difference regarding some of our results. For instance, whereas the relationship between the number of authors and citations is negative in marketing~\cite{Stremersch.2007}, it is positive in software engineering. While the presence of question marks in titles has a negative association with citations in biology and medicine~\cite{Jamali.2011}, we find the opposite in software engineering. Conversely, while acronyms drive citations in medicine~\cite{Jacques.2010}, they seem to inhibit them in software engineering.

\textbf{Implication \#4:} The effects are robust. Going beyond extant literature, we controlled for venue-year fixed effects, effectively only analyzing variation between articles within a venue in a given year. We additionally used robust standard errors to account for the non-independence of observations. Despite these advanced statistical controls, the results hold.

\subsection{Threats to validity}
\subsubsection{Internal validity}

There are several potential threats to validity. While we controlled for a host of potential confounders, we are hesitant to claim proof for causality. There might be other factors that we either deliberately do not include (because we focus on the early stage of the citation process) or cannot account for and which influence both our independent variables and drive citations. The quality and importance of each individual paper is such a potentially confounding factor. Unfortunately, controlling for paper quality is notoriously difficult and thus rarely done.
Initial work on research quality in empirical software engineering has attempted to operationalize quality as reported rigor and relevance, and manually scored both dimensions using rubrics that evaluate degrees of quality based on which information is disclosed in a focal article~\cite{Ivarsson.2011}. While this effort was influential and is certainly insightful, we did not apply it for four reasons. First, we propose that reviewers and editors at the different venues we study act as quality gatekeepers and apply reasonably consistent quality standards within each venue. We do thus not immediately see why fundamental article quality within a venue should vary strongly. Recall that any such venue effects are controlled for through the included dummy variables. Second, the quality measure captures reporting quality rather than necessarily actual quality of a focal article. While reporting quality may be important, it need not be indicative of the actual quality of the work~\cite{Ivarsson.2011}. Third, in prior work in the empirical software engineering literature~\cite{Molleri.2018}, relevance was not found to be related to citations. While one prior analysis suggests an association between rigor and citations, other features like article length positively co-varied with rigor and were not controlled for in the analysis, suggesting that the association may not be causal. Overall, it thus appears warranted to conclude that ``rigor, relevance and normalized citations are fairly orthogonal variables''~\cite[p. 1466]{Molleri.2018}. Fourth and critically, the measure is simply impractical to use in studies with samples as large as ours.

The texts we study do not represent the entire text of the scientific articles. While we welcome future studies that also analyze the full texts of articles, we consider ourselves in the good company of prior researchers who specifically studied titles and abstracts because these elements are the ones that set the ``hook'' and lure readers into reading an article in the first place~\cite{Mabe.2002}. In fact the abstract ``is often the only part of the article that will be read''~\cite{Pitkin.1999}, and abstracts and titles are crucial surrogate information~\cite{Janes.1991}.

Our study also has limitations stemming from the data we used. Specifically, WoS contains missing data for some variables in a number of observations. Combined with our multiple regression approach, this means that we had to disregard such observations. Prior research suffered from similar limitations~\cite{Mantyla.2019}.

\subsubsection{External validity}

Our analysis focuses on the top 20 venues in the field of empirical software engineering over a time span of 30 years. Any statements about the validity of our results beyond this context---say, regarding other disciplines, less influential or broader software engineering venues, and other points in time including the future---would thus be speculative.

Nevertheless, to make an attempt to empirically further assess generalizability, we re-ran our analysis over the entire sample of abstracts we obtained (\textit{N} = 255,857). Recall that we deemed this sample not fully reliable for our main analysis for several reasons, such as the inclusion of venues only marginally or not at all related to software engineering, and issues in harmonizing venue titles for such a large sample, making any venue-related fixed effects somewhat less reliable. To address the latter issue in the re-analysis, we performed a simple automated harmonization of venue names (e.g., removing numerals from conference names) to reduce the overall number of distinct venue names in the sample. Since the number of venues still remained in the thousands, however, we needed to simplify our analysis to make it computationally feasible. Specifically, we retained the venue and year dummies, but omitted interactions between the two from this analysis.

Our findings in this re-analysis on the entire sample largely corroborate our findings. Specifically, all coefficients that were significant in the main sample were also significant with the same sign in the entire sample (\textit{p} < 0.05), with the exception of the number of acronyms in the paper title, which is marginally significant in the opposite direction in the entire sample (\textit{p} < 0.1). Conversely, the two insignificant (length of title) or marginally significant (number of keywords) variables in the main sample were significant in the entire sample (\textit{p} < 0.001). Notably, the number of keywords displayed the initially hypothesized positive relationship with citations in the entire sample. Overall, the findings in the entire sample suggest that most of our findings have validity in a larger software engineering context.

Finally, as other researchers cautioned, using different bibliometric databases might lead to potentially different results. In fact, while this is a limitation, we also consider it a contribution of our paper that we use a different database than that used in the working paper that is closest to our work~\cite{Mantyla.2019}.

\section{Conclusion and Future Work}

Our results suggest that article metadata are related to citations. Authors might thus wish to purposefully consider how to phrase their titles and abstracts to attract a maximum number of citations~\cite{Grant.2013}. Due to the potential concerns about causality, however, we also encourage future researchers to build on our work and that of others and study drivers of citations in software engineering further. Particularly, studying ``later-stage'' characteristics of papers, such as the actual substantive content, appears fruitful. Comprehensive models, considering large arrays of characteristics simultaneously, as well as the use of comprehensive statistical control techniques like fixed effects, might be particularly useful in this quest.

\begin{acks}
We thank Michael A. Covington for providing us with CPIDR 5.
\end{acks}

\bibliographystyle{ACM-Reference-Format}
\bibliography{base}

\end{document}